\documentclass{article}
\usepackage{graphicx}
\usepackage{authblk}

\title{Boosting Transparency in Topological Josephson Junctions via Stencil Lithography}

\author[A,E,1,*]{Peter Sch\"uffelgen}
\author[A,E,1]{Daniel Rosenbach}
\author[B]{Chuan Li}
\author[A]{Tobias Schmitt}
\author[A]{Michael Schleenvoigt}
\author[A]{Abdur Rehman Jalil}
\author[A]{Jonas K\"olzer}
\author[C,E]{Meng Wang}
\author[A]{Benjamin Bennemann}
\author[A]{Umut Parlak}
\author[A]{Lidia Kibkalo}
\author[A]{Doris Meertens}
\author[A]{Martina Luysberg}
\author[A,E]{Gregor Mussler}
\author[B,D]{Alexander A. Golubov}
\author[B]{Alexander Brinkman}
\author[A,E]{Thomas Sch\"apers}
\author[A,E]{Detlev Gr\"utzmacher}
\affil[A]{Peter Gr\"unberg Institut, Forschungszentrum J\"ulich \& JARA J\"ulich-Aachen Research Alliance, D-52425 J\"ulich, Germany.}
\affil[B]{MESA+ Institute for Nanotechnology, University of Twente, 7500AE Enschede, The Netherlands.}
\affil[C]{State Key Laboratory of Functional Materials for Informatics, Shanghai Institute of Microsystem and Information Technology, Chinese Academy of Sciences, Shanghai 200050, China.}
\affil[D]{Moscow Institute of Physics and Technology, Dolgoprudny, Moscow District, Russia.}
\affil[E]{Helmholtz Virtual Institute for Topological Insulators (VITI), Forschungszentrum J\"ulich, D-52425 J\"ulich, Germany.}
\affil[1]{These authors contributed equally to this work}
\affil[*]{E-mail correspondence: p.schueffelgen@fz-juelich.de}
\date{} 

\begin{document}

\maketitle

\begin{abstract}
Hybrid devices comprised of topological insulator (TI) nanostructures in proximity to s-wave superconductors (SC) are expected to pave the way towards topological quantum computation. Fabrication under ultra-high vacuum conditions is necessary to attain high quality of TI-SC hybrid devices, because the physical surfaces of (Bi,Sb) based three-dimensional TIs suffer from degradation at ambient conditions. Here, we present an in-situ process, which allows to fabricate such hybrids by combining molecular beam epitaxy and stencil lithography under ultra-high vacuum conditions. As-prepared Josephson junctions were investigated by transmission electron microscopy and low temperature transport measurements. Results show highly transparent interfaces and large induced gaps. The Shapiro response of radio frequency measurements indicate the presence of gapless Andreev bound states, so-called Majorana bound states. These findings qualify (Bi$_{0.06}$Sb$_{0.94}$)$_2$Te$_3$ as material for Majorana devices, if combined with superconductive Nb by the presented in-situ technology. 
\end{abstract}

\section*{Introduction}
The concept of quantum computation promises to solve certain problems, which nowadays the most powerful supercomputers cannot$^1$. The Achilles heel of conventional (non-topological) concepts of quantum computing is the huge overhead needed for correcting quantum errors$^{2,3}$. Qubit layouts which use so-called Majorana modes have been predicted to compute fault-tolerantly with only little need of error correction$^{4,5,6,7,8}$. Localized Majorana modes form at the ends of quasi 1D topological insulator (TI) nanowires in proximity to an s-wave superconductor$^{9,10}$ (SC). One way to test whether or not a given combination of an s-wave SC and a 3D TI hosts Majorana modes is the Shapiro response of a Josephson junction (JJ).\\

The Josephson effect allows to conduct dissipationless supercurrents across a lateral junction of two close-by superconductive electrodes separated by a weak link of non-superconductive material. The supercurrent is mediated by so-called Andreev bound states (ABS), which effectively transport Cooper pairs across the weak link$^{11}$. JJs with weak links of 3D TI thin films have an additional transport channel, the so-called Majorana bound state (MBS)$^{12}$. The reason for this unconventional contribution lies in the nature of the topological surface states$^{13}$. Forbidden backscattering into the orthogonal state with opposite momentum $\vec{k}$ (Fig. 1a) makes some quasiparticles being perfectly transmitted into the SC (Fig. 1b)$^{14}$. MBS transfer only a single electron per cycle across the junction and reveal themselves by a suppression of odd Shapiro steps in low temperature transport experiments under radio frequency (RF) radiation, due to their 4$\pi$-periodic energy phase dispersion (Fig. 1c)$^{15}$.\\

\begin{figure}[ht]
\includegraphics[keepaspectratio=true,width=\textwidth]{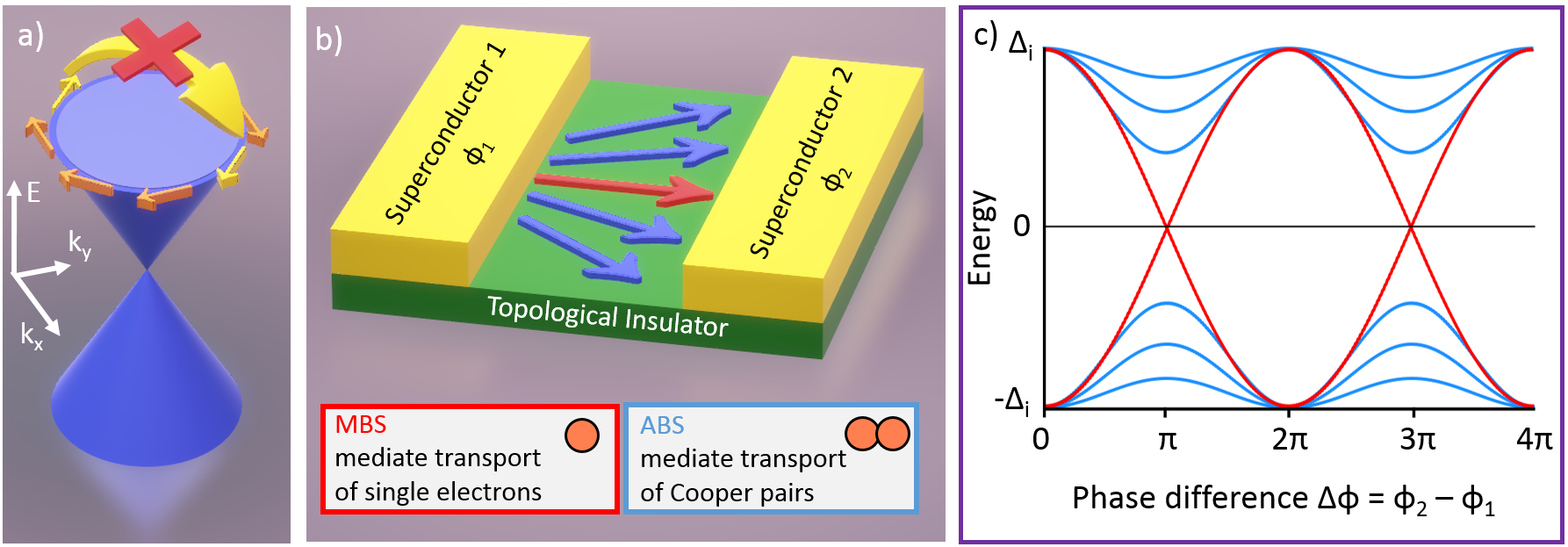}
\caption{Forbidden backscattering in topological surfaces and its consequences for transport in topological Josephson junctions: a) The Dirac cone describes the energy dispersion of the 2D surface states in 3D TIs.  The spins of the Dirac surface states are locked to their momenta. Forbidden backscattering into the orthogonal surface state with opposite spin and $\vec{k}$ (yellow arrows) has direct implications for transport in topological Josephson junctions. b) Bound states, which impinge at the superconducting electrode perpendicular (red arrow), are perfectly transmitted, due to the forbidden backscattering. c) Therefore, Majorana bound states (MBS) are not gapped (red), in contrast to non-perpendicular (Andreev) bound states (blue). Because of the zero energy crossing, MBS show a 4$\pi$-periodic energy-phase dispersion ($E(\Delta\phi)$). MBS mediate transport of single quasiparticles across the junction$^{25}$. ABS on the other hand mediate transport of Cooper pairs.}
\end{figure}

The first signatures of transport mediated by MBS in 3D TI JJs were reported on Nb-HgTe-Nb junctions$^{16}$. Although similar experiments were carried out on Bi-based topological JJs$^{17,18}$, 4$\pi$-periodic signatures have only been observed from bulk states in Dirac semimetals$^{19}$, but not from surface states of TIs. Different to HgTe the surface oxidation$^{20,21}$ and reactions with water molecules$^{22}$ can lead to additional non-topological states at the surface of Bi-based TIs. These superimpose locally with the spin-momentum locked TI states and therefore backscattering of the perpendicular mode is possible again, destroying the 4$\pi$-periodic channel. The surface degradation, however, can be avoided by supplying a protective capping layer on top of the topological surface in-situ, right after the growth of TI thin films$^{23,24}$. A significant disadvantage, which comes along with in-situ capping, is the difficulty in gentle but thorough removal of the capping layer for deposition of highly transparent superconductive electrodes$^{25}$. By depositing first the electrodes and then the capping, both under ultra-high vacuum conditions, surface protection is guaranteed and highly transparent interfaces in between electrodes and TI are obtained, as demonstrated in this work.\\
\newpage
Here we present a fabrication process, which allows one to prepare such junctions with electrode distances down to few tens of nm. The technique can be adapted to arbitrary layouts and opens the door towards more complex Majorana devices based. Especially, when combined with other bottom-up techniques like selective area growth$^{26,27,28,29,30}$ our process paves the way towards networks of in-situ grown Majorana devices for proposed topological qubit layouts$^{5}$.

\section*{Stencil Lithography of in-situ grown Josephson Devices}
For stencil fabrication the surface of the Si-substrate was covered by a thin SiO$_2$ interlayer, separating the subsequently deposited and defined Si$_3$N$_4$ hard mask from the Si surface. Both, the 300\,nm SiO$_2$ and the 100\,nm Si$_3$N$_4$ were globally deposited in a Centrotherm LPCVD System E1200 R\&D furnace on a 4" n-type Si(111) substrate ($<2000$ $\Omega\cdot$cm) (Fig. 2a), right after cleaning the Si surface wet chemically by the RCA HF-last procedure$^{31}$. Subsequently, the top Si$_3$N$_4$ layer was patterned via electron beam lithography and reactive ion etching. In this manner, holes for subsequent electrode deposition were defined, which are only 50-100\,nm apart from each other at their narrowest point (Fig. 2b). A dip in 1\% hydrofluoric acid removes the exposed SiO$_2$ isotropically. Due to the small electrode distance, the SiO$_2$ underneath the Si$_3$N$_4$ is fully removed, releasing a freestanding Si$_3$N$_4$ bridge at a height of 300\,nm above the substrate (Fig. 2c). Additionally, the dip in 1\% hydrofluoric acid passivates the surface dangling bonds of the Si(111) substrate during ex-situ transfer to the MBE growth chamber$^{31}$. The hydrogen passivation is removed by keeping the substrate at 700\,$^{\circ}$C for 20 minutes under ultra-high vacuum conditions. When the substrate temperature stabilized at the growth temperature of 280\,$^{\circ}$C, a (Bi$_{0.06}$Sb$_{0.94}$)$_2$Te$_3$ thin film was grown in a Te-overpressure regime (T$_{Te}$ = 330\,$^{\circ}$C, T$_{Sb}$  = 450\,$^{\circ}$C and T$_{Bi}$  = 450\,$^{\circ}$C), following the recipe of Kellner et al$^{32}$. At 280\,$^{\circ}$C the TI grows selectively on the Si(111) surface, while no material is deposited on top of the Si$_3$N$_4$ mask. The molecular beams impinge at the substrate under a polar angle of 32.5$^{\circ}$ - defined by the geometry of the MBE system, and therefore the steady substrate rotation (10 min$^{-1}$) made the TI thin film grow homogenously underneath the freestanding bridge (Fig. 2d).\\

\begin{figure}[ht]
\includegraphics[keepaspectratio=true,width=\textwidth]{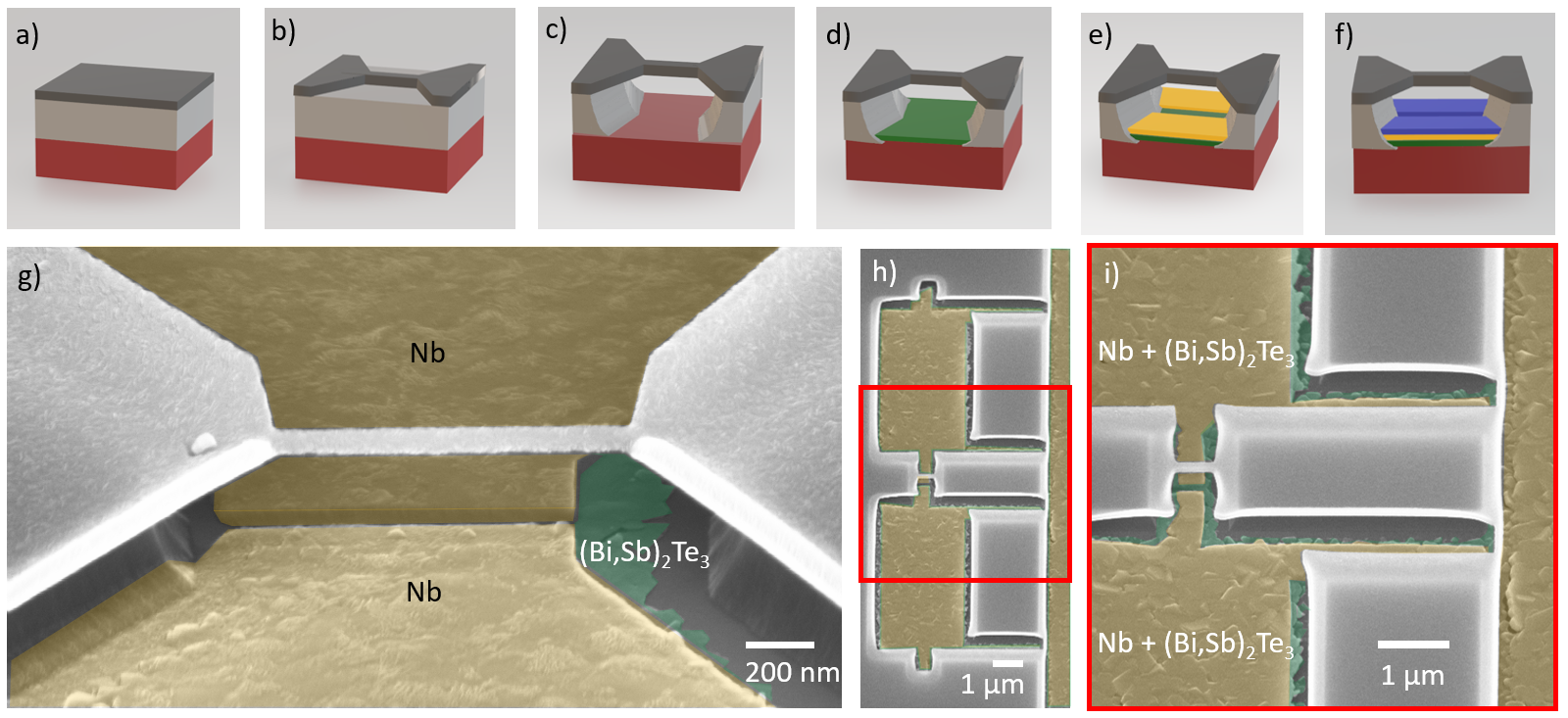}
\caption{In-situ fabricated Josephson junctions: a) Stack of 525\,$\mu$m Si (red), 300\,nm SiO$_2$ (white) and 100\,nm Si$_3$N$_4$ (grey). b) Defining the Si$_3$N$_4$ mask via electron beam lithography and reactive ion etching. c) Release of the mask by hydrofluoric acid etching. d) Growth (with substrate rotation) of TI (green). e) Deposition of superconductive electrodes (yellow) via stencil lithography (without rotation). Note that superconductive and dielectric thin films are also deposited on top of the mask, but not shown in the scheme for clarity. f) Deposition (with rotation) of dielectric cap-layer (blue). g) False colour scanning electron micrograph of an as-prepared JJ: yellow = Nb-electrodes, green = (Bi,Sb)$_2$Te$_3$ thin film, the dielectric Al$_2$O$_3$ capping layer is not visible in the scanning electron micrograph. h) Towards more complex hybrid devices: Multi array comprised of five Josephson junctions. i) Close-up of (h).}
\end{figure}

For deposition of the superconductive electrodes, the sample has been transferred via a portable vacuum chamber to a metal MBE-chamber, continuously keeping the pressure below 10$^{-8}$\,mbar. During the deposition of 50\,nm Nb the substrate rotation was turned off and the orientation of the Si$_3$N$_4$ bridge was aligned to the position of the Nb pocket with an accuracy of $\pm$1$^{\circ}$ (azimuth angle). The polar angle with respect to the surface normal was again 32.5$^{\circ}$ in this chamber. The TI thin film is covered completely by Nb, except for the narrow area, which has been shaded by the free-standing nano-bridge. This shaded area forms the weak link in between the two superconductive electrodes (Fig. 2e). For the capping of the TI weak link, the substrate rotation has been set to 10 min$^{-1}$ and 5\,nm of Al$_2$O$_3$ were deposited, covering the whole sample (Fig. 2f). After those steps the device is ready for testing and no further ex-situ fabrication is necessary. Fig. 2g) shows a perspective false colour scanning electron micrograph of a single JJ. The thin Al$_2$O$_3$ is not observable in the scanning electron micrograph, but clearly visible in the transmission electron micrograph (Fig. 3a). To demonstrate scalability, more complex layouts were fabricated; an as-prepared array of Josephson devices is shown in Fig. 2h) and i). Note that in all micrographs the Si$_3$N$_4$ masks are still in contact with the substrate via the SiO$_2$ interlayer and are as well covered with Nb and Al$_2$O$_3$, since those materials were deposited all over the sample.\\

\begin{figure}[ht]
\includegraphics[keepaspectratio=true,width=\textwidth]{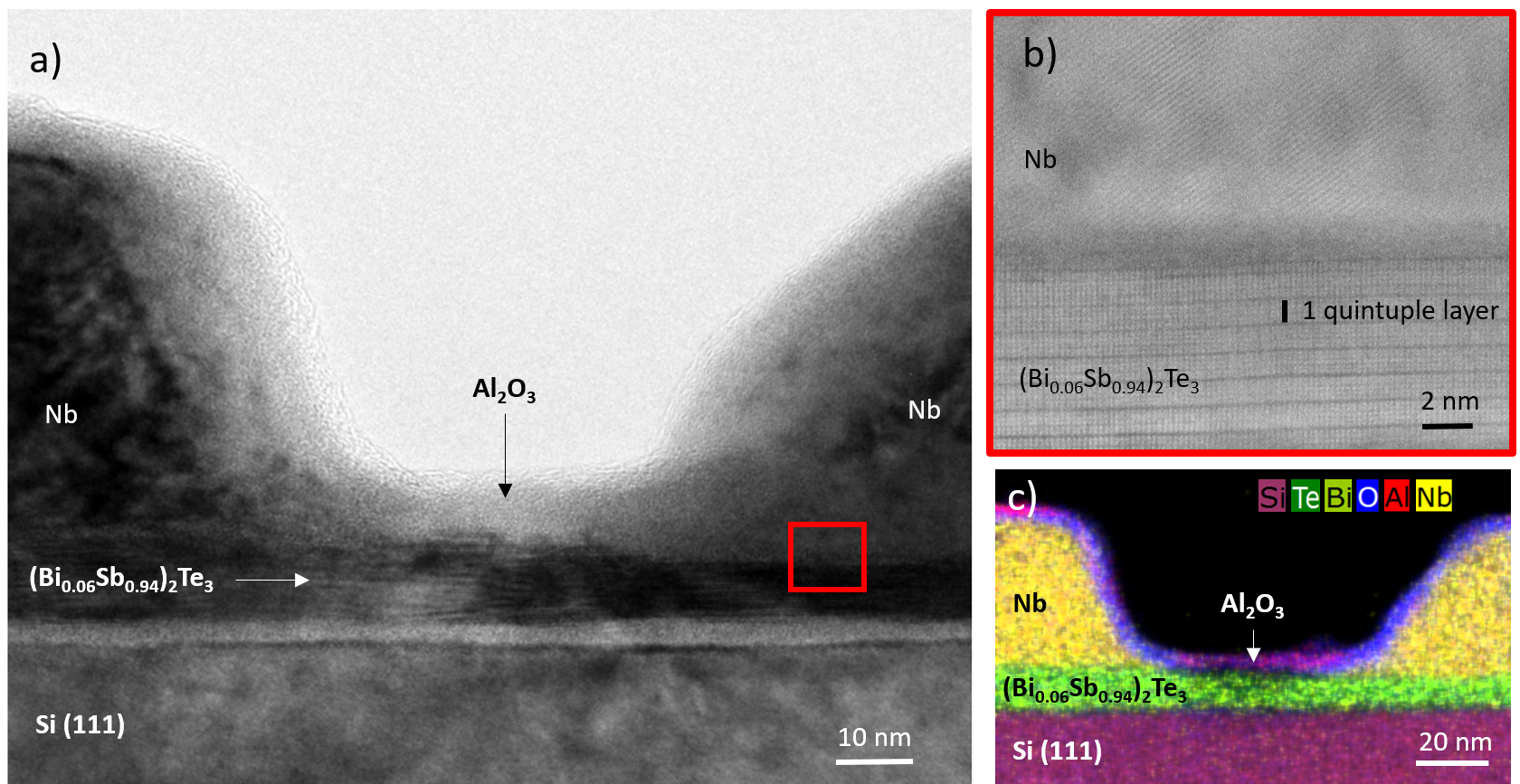}
\caption{Cross section of a single topological Josephson junction:  a) TEM bright field image displaying an overview of the junction. The layered structure of the rhombohedral (Bi,Sb)$_2$Te$_3$ crystal and the crystallinity of the Nb electrode is clearly visible in the high angle annular dark field image in Fig. 3b (obtained from the red framed region in (a)). In between the Nb electrodes, a 5\,nm thick Al$_2$O$_3$ layer protects the weak link from degradation. c) EDX map of Fig. 3a indicates the distribution of individual elements.}
\end{figure}

 A single JJ is seen in cross section in the transmission electron microscope (TEM) bright field image in Fig. 3a). The crystalline nature of both, the TI and the Nb electrodes is revealed in the high angle annular dark field image (Fig. 3b). Lattice fringes appear in the Nb film, whereas the rhombohedral (Bi,Sb)$_2T$e$_3$ crystal is identified by the quintuple layers$^{33}$, separated by van der Waals bonds (dark horizontal lines in the lower part of Fig. 3b). Fig. 3c) displays an energy-dispersive X-ray spectroscopy (EDX) map, where the contributions of the different elements are displayed in different colours. The Nb-(Bi,Sb)$_2$Te$_3$ interface exhibits an interdiffusion layer of about 1\,nm which can be seen in Fig. 3b). The distance of the two superconductive Nb electrodes is about 60\,nm for this junction. The area between the electrodes is fully covered with the desired Al$_2$O$_3$ capping layer, protecting the surface of the TI crystal from exposure to ambient conditions when moving the sample to air.\\

\section*{Low temperature transport experiments}
We measured five (Bi$_{0.06}$Sb$_{0.94}$)$_2$Te$_3$ JJs in a dilution refrigerator. All devices have been fabricated via the stencil technique presented above of materials shown in Fig. 3. The JJs varied only in width and length of the weak link. Each device included interconnect lines (shown in the insert of Fig. 4a) permitting standard 4-point transport measurements: one pair to supply AC+DC current ($I$) across the junction and a second pair for measuring the voltage ($V$) and differential resistance ($dV/dI$) across the weak link. The normal resistances across the weak links scale accordingly to the junctions’ dimensions (see supplementary). At low temperatures, a dissipationless supercurrent ($V =0$) is observed for small $I$ in all junctions. In Fig. 4a the $I$-$V$  curve of a 620\,nm wide and 30\,nm long junction at 12\,mK is shown. Depending on the sweep direction, the retrapping current at which the voltage returns to zero is $I_R = 2.2\,\mu$A, much smaller than the critical switching current $I_C =6.2\,\mu$A. This large hysteresis can be explained by self-heating effects or capacitance$^{34}$ and has been observed before in topological JJs$^{35}$. At higher temperatures, $I_C$ is decreased and above 2.2\,K we measure non-hysteretic $I$-$V$ curves (Fig. 4b).\\

\begin{figure}[ht]
\includegraphics[keepaspectratio=true,width=\textwidth]{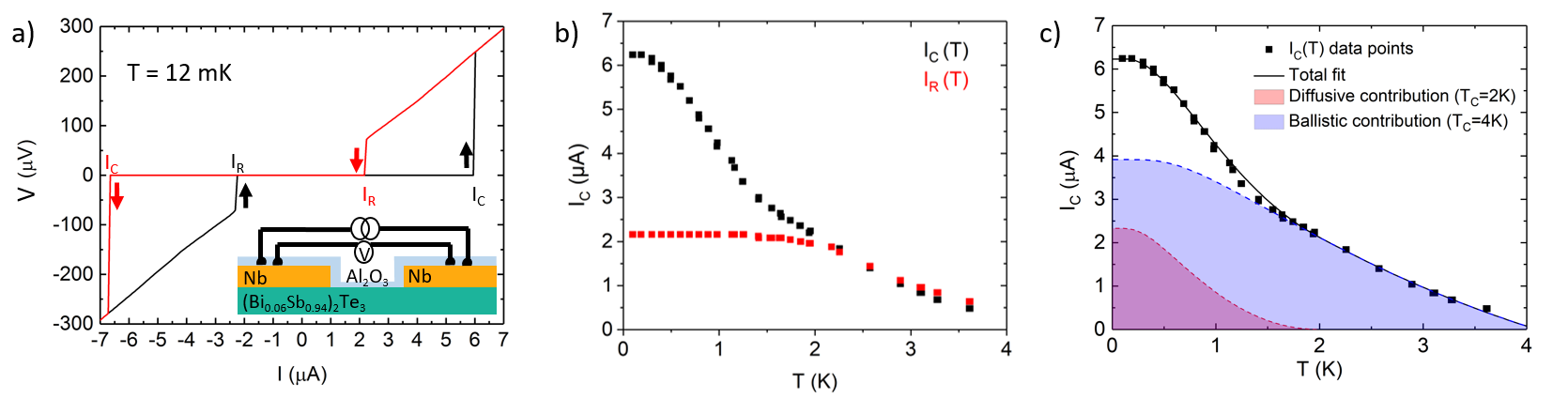}
\caption{DC transport of a topological JJ and its temperature dependency. a) $I$-$V$ characteristics at 12\,mK for a 620\,nm wide and 30\,nm long junction. b) $I_C (T)$ and $I_R (T)$ of (a). For temperatures above $2.2\,$K the supercurrent is symmetric ($I_c = I_R$) c) $I_C (T)$ in (b) is comprised of ballistic and diffusive contributions. For temperatures above 2.2\,K the supercurrent is solely carried by ballistic contributions.}
\end{figure}

The slope of $I_C (T)$ shows a kink at 1.4\,K, because not only the TI surface, but also bulk modes contribute to the supercurrent$^{17}$. The Fermi level of the upper surface can be tuned very accurately$^{36}$ and has been found to be in the bulk band gap for (Bi$_{0.06}$Sb$_{0.94}$)$_2$Te$_3$$^{33}$. However, normal transport in these films indicates that the Fermi level in the bulk crosses the bulk valence bands (see supplementary) due to band bending effects at the upper surface$^{37,38}$. The bulk contributions in our junction have a smaller mean free path and therefore form a diffusive channel next to the ballistic channel carried by the surface states. In superconducting transport, the ballistic modes are favored for charge transfer$^{17}$. In Fig. 4c) the temperature dependence of both contributions is fitted. At the base temperature, the contribution of surface modes is significantly larger than that of bulk modes ($\sim 2.3\,\mu$A), opposite than in normal state transport in Hall bar devices with electrodes separated tens to hundreds of $\mu$m$^{23,39}$. Furthermore, the ballistic modes turn out to have a higher $T_c$ than the bulk modes. While the diffusive channel has a critical temperature of only 2\,K, the ballistic modes survive up to 4\,K. Thus above 2\,K the supercurrent is solely carried by ballistic surface modes. For lower temperatures more and more diffusive modes contribute. At base temperature we measure a normal resistance  of $R_N\geq50\,\Omega$ and obtain an $I_CR_N \geq310\,\mu$V for our sample (Fig 4). Compared to $I_CR_N$ products cited in literature$^{19}$, our value is rather high, indicating a strong superconducting proximity effect and a large induced superconducting gap. These properties are the consequence of interface transparency, which is an objective measure of the probability of successful charge transport via ABS across the junction. An interface transparency of 1 has a 100\% probability for Andreev-reflected quasiparticles and is therefore defined as perfectly transparent. By fitting the ballistic contribution of the supercurrent with the Eilenberger equation we obtain a value of 0.95 for the transparency of our junction (for more details on the theory fit to $I_c (T)$ data, see supplementary material). Note, that this transparency relates to the in-plane boundary between proximitized TI and non-proximitized TI. The perpendicular proximity effect from Nb into TI is also strong, as reflected by the large critical temperature of the proximity induced gap (4\,K). We attribute this to the in-situ deposition of Nb. The in-situ capping on the other hand assures that the ballistic contribution of the supercurrent is solely carried by Dirac states without contributions from additional trivial (= non-topological) states at degraded surfaces. Shapiro step measurements above 2\,K allow to disentangle the relative contributions of ABS and MBS to the ballistic supercurrent. 

\section*{Shapiro response}
In JJs with trivial weak links, Cooper pair transport across the junction is mediated by ABS$^{11}$. For topological junctions more complex models have been proposed$^{15}$, which consider MBS as an additional transport channel. While ABS are 2$\pi$-periodic (Fig. 1c) and effectively transport correlated pairs across the junction, 4$\pi$-periodic MBS mediate transport of single quasiparticles$^{25}$. For detecting signatures of MBS in transport experiments, it is beneficial to reduce the number of 2$\pi$-periodic ABS. The stoichiometric compound (Bi$_{0.06}$Sb$_{0.94}$)$_2$Te$_3$ has its Fermi level ($E_F$) at the upper surface as close as 2$\pm$7\,meV to the Dirac point ($E_D$), i.e. it has a small Fermi vector of $\left|\vec{k_F}\right| = \frac{v_D\hbar}{E_F-E_D}<0.04$\,\r{A}$^{-1}$. With v$_D\geq3.9\pm0.2$ $10^5$\,m/s being the Dirac velocity$^{32}$. A small Fermi vector reduces the number of surface ABS in a JJ. The number of 2$\pi$-periodic modes (M) in a junction of width $w=620\,$nm can be estimated$^{40}$ to $M=\frac{\left|\vec{k_F}\right|\cdot w}{\pi}<75$. It is known that the deposition of electrodes can result in doping of the weak link area. Although the fine-tuned Fermi level might have shifted in our JJs due to Nb deposition, we still assume a reduced number of modes compared to JJs of binary compounds. Therefore, the relative contribution of possible 4$\pi$-periodic MBS should still be enhanced in JJs with a (Bi$_{0.06}$Sb$_{0.94}$)$_2$Te$_3$ weak link. Due to the AC Josephson effect, RF irradiation of a JJ results in voltage steps at constant height in the $I$-$V$ curve. Measuring those Shapiro steps allows one to investigate the ratio of MBS to ABS contributing to the total critical current I$_C$=I$_{4\pi}+I_{2\pi}$. When JJs are irradiated with an RF signal, the corresponding photons are absorbed within the weak link area, increasing the energy of the charge carriers within. The excited quasiparticles can only enter the second electrode, if its chemical potential $\mu_2$ is shifted by $V=\mu_2-\mu_1=n\cdot \frac{hf}{q}$. Here, $q$ resembles the transmitted charge, $h$ is the Planck constant, $n$ is an integer number, $f$ the frequency of the RF-signal and $V$ the voltage across the junction. For Cooper pairs $q$ is 2e. Under RF irradiation steps evolve at constant voltages $n\cdot V_{ABS}=n\cdot \frac{hf}{2e}$ within the $I$-$V$ characteristic and corresponding dips appear in $dV/dI(V)$ plots (Fig. 5a)$^{41}$. For MBS, however, the 2e charge reduces to 1e, i.e. the odd steps disappear and only even steps should evolve for pure transport via MBS (Fig. 5b). For 2D TIs this has indeed been observed and the disappearance of odd steps up to the ninth step has been reported$^{42}$. For 3D TIs of the same material, only the first step disappeared$^{16}$. According to Dominguez et al.$^{15}$, this is due to the additional 2$\pi$-periodic ABS, which carry most of the supercurrent in JJs with 3D TI interlayer. Although the final proof of Majorana modes can be only furnished by unveiling their non-abelian nature, a missing first Shapiro step is considered to be already an indicator of those exotic modes in topological JJs.\\

\begin{figure}[ht]
\includegraphics[keepaspectratio=true,width=\textwidth]{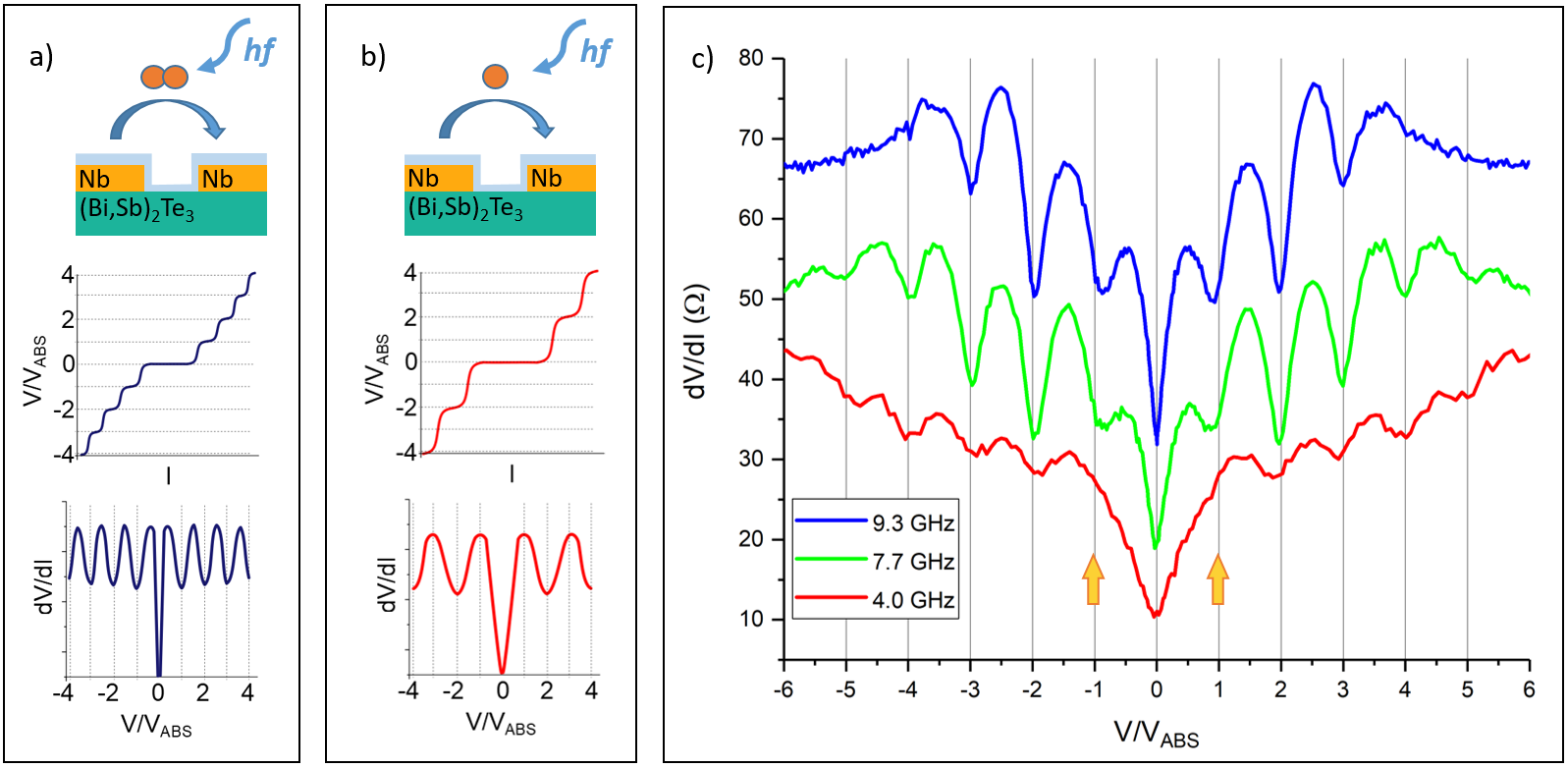}
\caption{a) Schematic Shapiro response of ABS: Cooper pairs, which absorb a photon with energy hf, while passing the weak link, trigger integer Shapiro steps of constant voltages $n\cdot V_{ABS}=n\cdot \frac{hf}{2e}$ within the $I$-$V$ curve and corresponding dips in $dV/dI$. b) Schematic Shapiro response of MBS: The period of steps is doubled compared to (a). The absorption of the photon by a single electron leads to Shapiro steps at integer multiples of $V_{MBS}=\frac{hf}{e}$, which is twice the step height, since  $n\cdot V_{MBS}=2n\cdot V_{ABS}$. c) Measured Shapiro response of our junction at $T=3.2\,$K: For high frequencies (9.3\,GHz) the transport is dominated by ABS, resulting in a clear 2$\pi$-periodic response with all integer steps visible. For lower frequencies (7.7\,GHz) the first dip is attenuated and for 4.0\,GHz the first step is fully suppressed (yellow arrows), suggesting that MBS are contributing to the total current. Note that the curves are vertically shifted for clarity. }
\end{figure}

We carried out Shapiro measurements on our in-situ fabricated Nb-(Bi$_{0.06}$ Sb$_{0.94}$)$_2$Te$_3$-Nb junctions at 3.2\,K to rule out missing steps due to hysteresis and to assure transport is solely carried by surface states (see supplementary). Because of thermal broadening, we plotted the differential resistance, where Shapiro steps are more pronounced at such high temperatures (Fig. 5c). For an RF frequency of 9.3\,GHz Shapiro steps evolve at integer multiples of the characteristic voltage $V_{ABS}=n\cdot \frac{h\cdot 9.3GHz}{2e}$. The first three steps are clearly visible as local minima in $dV/dI$. By reducing the frequency of the RF-signal to 7.7\,GHz the first step gets attenuated. For 4.0\,GHz we observe a fully suppressed first Shapiro step (yellow arrows in Fig. 5c). Although, Shapiro step measurements have been carried out on (Bi,Sb)$_2$Te$_3$ JJs before, the missing first step was not observed$^{17,18}$. Similar as the capping layer, the high interface transparency, obtained by in-situ fabrication, plays a crucial role for entering the regime in which 4$\pi$-periodic signatures become observable. Each contribution, $I_{4\pi}$ and $I_{2\pi}$ has a characteristic Josephson-frequency: $f_{4\pi} = I_{4\pi} R_N  \frac{e}{h}$ and $f_{2\pi}=I_{2\pi} R_N  \frac{2e}{h}$. Following Wiedenmann et al.$^{16}$, once the external frequency is on the order of one of these frequencies, the corresponding transport mechanism is favored. The suppression of odd Shapiro steps becomes observable only when the driving frequency becomes smaller than f$_{4\pi}$. Hence, due to the full suppression of the first step at 4.0\,GHz, the 4$\pi$-periodic contribution of the supercurrent could be estimated at around 400\,nA. Interfaces with lower transparency will have lower transition frequencies, which are hardly accessible in experiments.

\section*{Conclusion}
To conclude, we showed how to fabricate in-situ SC-TI hybrid devices by combining molecular beam epitaxy and stencil lithography. Both, the capping layer and highly transparent interfaces are necessary to observe signatures of MBS in (Bi,Sb)-based topological insulator JJs. Transmission electron micrographs confirm crystallinity of the electrodes and full capping of the weak link. The interface transparency has been estimated via an Eilenberger fit showing that it approaches 1. After disentangling bulk from surface contributions to the total critical current, we estimated the current mediated via MBS to be of the order of 400\,nA. The process presented here is suitable for fabrication of hybrid devices with a large throughput and a high interface quality. Attaching the stencil mask at a well-defined distance of 300\,nm from the substrate surface enables to reproducibly fabricate layouts with nm precision. Our results demonstrate the usefulness of in-situ techniques to fabricate more complex devices for further investigating Majorana physics in SC-TI hybrids. Especially, when combined with other in-situ bottom up techniques like selective area growth$^{26,27,28,29,30}$ , this method offers a scalable approach to fabricate proposed Majorana qubits and might pave the way towards fault-tolerant topological quantum computation.

\section*{Methods}
Structural investigations on the atomic scale have been performed with an aberration-corrected high-resolution STEM (FEI Titan 80–300)$^{43}$ equipped with a Super-X detector. This makes possible for simultaneous acquisition of high-resolution, high angle annual dark field images and energy dispersive X-ray spectra. Hence, structure and composition of crosssectional specimen can be determined. Specimens have been prepared with a Helios NanoLab 400S$^{44}$  by focused ion beam (FIB) milling, using first 30\,keV Ga ions followed by a 5\,keV final treatment. Ar ion milling with the NanoMill operated at 900\,V and subsequently at 500\,V was employed to reduce the surface damage introduced by FIB.

\section*{Author Contributions}
P.Sch\"u. and Da.R. contributed equally to this work. P.Sch\"u, Da.R., To.Schm. and M.Schl. fabricated the substrates in the cleanroom. P.Sch\"u., M.Schl., A.J. M.W. and G.M. grew the topological insulator thin films via MBE. B.B. grew the superconducting Nb. U.P. capped the sample with stoichiometric Al2O3. P.Sch\"u., Da.R., C.L. and To. Schm. performed the electrical transport measurements on Josephson devices. Da.R. and J.K. investigated the magneto-transport on Hall bars. L.K., Do.M. and Ma.Lu. prepared the FIB lamellas and performed HR STEM measurements. A.Go. and A.Br. carried out the Eilenberger and Usadel-fitting. P.Sch\"u., Da.R. and A.Br. wrote the paper with contributions from all coauthors. P. Sch\"u. initiated the project, which was supervised by A.Br., Th.Sch\"a. and D.Gr.\\

All authors have given approval to the final version of the manuscript.

\section*{Acknowledgements}
A. Braginski, F. Hassler, N. Tas and E. Berenschot are gratefully acknowledged for enlightening discussions. M. Geitner, K. Deussen and St. Trellenkamp are acknowledged for the deposition and patterning of the Si$_3$N$_4$ and SiO$_2$ layers. This work is supported by the German Science Foundation (DFG) under the priority program SPP1666 “Topological Insulators”, as well as by the Helmholtz Association via the “Virtual Institute for Topological Insulators” (VITI).

\section*{References}
1. Reiher, M., Wiebe, N., Svore, K. M., Wecker, D. \& Troyer, M. Elucidating reaction mechanisms on quantum computers, \textit{PNAS} \textbf{114}, 7555–7560 (2017).\\
2. Shor, P. W. Scheme for reducing decoherence in quantum computer memory, \textit{Phys. Rev. A} \textbf{52}, 2493–2496 (1995).\\
3. Kelly, J. et al. State preservation by repetitive error detection in a superconducting quantum circuit. \textit{Nat.} \textbf{519}:66 (2015).\\
4. Kitaev, A. Y. Unpaired Majorana fermions in quantum wires. \textit{Phys. Usp.} \textbf{44}, 131-136 (2001).\\
5. van Heck, B., Akhmerov, A. R., Hassler, F., Burrello, M. \& Beenakker, C. W. J. Coulomb-assisted braiding of Majorana fermions in a Josephson junction array. \textit{New J. Phys.} \textbf{14}, 035019 (2012).\\
6. Hyart, T., van Heck, B., Fulga, I. C., Burrello, M., Akhmerov, A. R. \& Beenakker, C. W. J. Flux-controlled quantum computation with Majorana fermions. \textit{Phys. Rev. B} \textbf{88}, 035121 (2013).\\
7. Fulga, I. C., van Heck, B., Burrello, M. \& Hyart, T. Effects of disorder on coulomb-assisted braiding of Majorana zero modes. \textit{Phys. Rev. B} \textbf{88},155435 (2013).\\
8. Pedrocchi, F. L. \& DiVincenzo, D. P. Majorana braiding with thermal noise. \textit{Phys. Rev. Lett.} \textbf{115}, 120402 (2015).\\
9. Manousakis, J., Altland, A., Bagrets, D., Egger, R. \& Ando, Y. Majorana qubits in a topological insulator nanoribbon architecture. \textit{Phys. Rev. B} \textbf{95}, 165424 (2017).\\
10. Cook,  A. \& Franz, M. Majorana fermions in a topological-insulator nanowire proximity-coupled to an s-wave superconductor. \textit{Phys. Rev. B} \textbf{84},201105 (2011).\\
11. Golubov, A. A., Kupriyanov, M. Y. \& Il’ichev, E. The current-phase relation in Josephson junctions. \textit{Rev. Mod. Phys.} \textbf{76}, 411–469 (2004).\\
12. Fu, L. \& Kane, C. L. Superconducting proximity effect and Majorana fermions at the surface of a topological insulator. \textit{Phys. Rev. Lett.} \textbf{100},096407 (2008).\\
13. Hasan, M. Z. \& Moore, J. E. Three-dimensional topological insulators.\textit{Annu. Rev. Condens. Matter Phys.} \textbf{2}, 55–78 (2011).\\
14. Dominguez, F. Hassler, F. \& Platero, G. Dynamical detection of Majorana fermions in current-biased nanowires. \textit{Phys. Rev. B} \textbf{86}, 140503 (2012).\\
15. Dominguez, F. et al. Josephson junction dynamics in the presence of 2$\pi$- and 4$\pi$-periodic supercurrents. \textit{Phys. Rev. B} \textbf{95}, 195430 (2017).\\
16. Wiedenmann, J. et al. 4$\pi$-periodic Josephson supercurrent in HgTe-based topological Josephson junctions. \textit{Nat. Com.} \textbf{7}, 10303 (2015).\\
17. Veldhorst, M. et al. Josephson supercurrent through a topological insulator surface state. \textit{Nat. Mater.} \textbf{11}, 417 (2012).\\
18. Galletti, L. et al. Influence of topological edge states on the properties of Al/Bi$_2$Se$_3$/Al hybrid Josephson devices. \textit{Phys. Rev. B} \textbf{89},134512 (2014).\\
19. Li, C. et al. 4$\pi$ periodic Andreev bound states in a Dirac semimetal. \textit{ArXiv e-prints}, arXiv:1707.03154 (2017).\\
20. Ngabonziza, P. et al. In situ spectroscopy of intrinsic Bi$_2$Te$_3$ topological insulator thin films and impact of extrinsic defects. \textit{Phys. Rev. B} \textbf{92}, 035405 (2015).\\
21. Thomas, C. R. et al. Surface oxidation of Bi$_2$(Te,Se)$_3$ topological insulators depends on cleavage accuracy. \textit{Chem. Mater.} \textbf{28}, 35–39, (2016).\\
22. Benia, H. M., Lin, C., Kern, K. \& Ast, C. R. Reactive chemical doping of Bi$_2$Se$_3$ topological insulator. \textit{Phys. Rev. Lett.} \textbf{107}, 177602 (2011).\\
23. Lang, L. et al. Revelation of topological surface states in Bi$_2$Se$_3$ thin films by in situ Al passivation. \textit{ACS Nano} \textbf{6}, 295–302 (2012).\\
24. Ngabonziza, P., Stehno, M. P., Myoren, H., Neumann, V. A., Koster, G. \& Brinkman, A. Gate-tunable transport properties of in situ capped Bi$_2$Te$_3$topological insulator thin films. \textit{Adv. El. Mater.} \textbf{2}, 1600157 (2016).\\
25. Sch\"uffelgen, P. et al. Stencil lithography of superconducting contacts on MBE-grown topological insulator thin films. \textit{J. Cryst. Growth} \textbf{477}, 183–187 (2017)\\
26. Fu, L., Electron Teleportation via Majorana Bound States in a Mesoscopic Superconductor. \textit{Phys. Rev. Lett.} \textbf{104}, 056402 (2010).\\
27. Gooth, J. et al. Ballistic One-Dimensional InAs Nanowire Cross-Junction Interconnects. \textit{Nano Lett.} \textbf{17}, 2596-2602 (2017).\\
28. Friedl, M. et al. Template-Assisted Scalable Nanowire Networks. \textit{Nano Lett.} \textbf{18}, 2666-2671 (2018).\\
29. Krizek, F. et al. Field effect enhancement in buffered quantum nanowire networks. \textit{ArXiv e-prints}, arXiv:1802.07808v2 (2018).\\
30. Kampmeier, J. et al. Selective area growth of Bi2Te3 and Sb2Te3 topological insulator thin films. \textit{J. Cryst. Growth} \textbf{443}, 38-42 (2016).\\
31. Kampmeier, J., Borisova, S., Plucinski, L., Luysberg, M., Mussler, G. \& Grützmacher, D. Suppressing twin domains in molecular beam epitaxy grown Bi$_2$Te$_3$ topological insulator thin films. \textit{Cryst. Growth Des.} \textbf{15}, 390–394 (2015).\\
32. Kellner, J. et al. Tuning the dirac point to the Fermi level in the ternary topological insulator (Bi$_{1-x}$Sb$_x$)$_2$Te$_3$. \textit{Appl. Phys. Lett.}
\textbf{107}, 251603 (2015).\\
33. Cava, R. J., Ji, H., Fuccillo, M. K., Gibson, Q. D. \& Hor, Y. S. Crystal structure and chemistry of topological insulators. \textit{J. Mater. Chem. C} \textbf{1}, 3176–3189 (2013).\\
34. Courtois, H., Meschke, M., Peltonen, J. T. \& Pekola, J. P. Origin of hysteresis in a proximity Josephson junction. \textit{Phys. Rev. Lett.} \textbf{101}, 067002 (2008).\\
35. Oostinga, J. B. et al. Josephson supercurrent through the topological surface states of strained bulk HgTe. \textit{Phys. Rev. X} \textbf{3}, 021007 (2013).\\
36. Weyrich, C. et al. Growth, characterization, and transport properties of ternary (Bi$_{1-x}$Sb$_x$)$_2$Te$_3$ topological insulator layers. \textit{J. Phys. Condens. Matter} \textbf{28}, 495501 (2016).\\
37. Brahlek, M., Koirala, N., Bansal, N. \& Oh, S. Transport properties of topological insulators: Band bending, bulk metal-to-insulator transition and weak anti-localization. \textit{Solid State Commun.}\textbf{15-216}, 54–62 (2015).\\
38. ViolBarbosa, C. E. et al. Direct observation of band bending in the topological insulator Bi$_2$Se$_3$. \textit{Phys. Rev. B} \textbf{88}, 195128 (2013).\\
39. Weyrich, C. et al. Magnetoresistance oscillations in MBE-grown Sb$_2$Te$_3$ thin films. \textit{Appl. Phys. Lett.} \textbf{110}, 092104 (2017).\\
40. Beenakker, C. W. J. \& van Houten, H. Quantum transport in semiconductor nanostructures. \textit{Solid State Physics} \textbf{44}, 1–228 (1991).\\
41. Shapiro, S. Josephson currents in superconducting tunneling: The effect of microwaves and other observations. \textit{Phys. Rev. Lett.} \textbf{11}, 80–82 (1963).\\
42. Bocquillon, E. et al. Gapless Andreev bound states in the quantum spin hall insulator HgTe. \textit{Nat. Nanotechnol} \textbf{12}, 137 (2016).\\
43. Kovacs, A., Schierholz, R. \& Tillmann, K. Fei titan g2 80-200 crewley. \textit{JLSRF}, \textbf{2}, 43 (2016).\\
44. Meertens, D., Kruth, M., \& Tillmann, K. FEI Helios NanoLab 400S FIB-SEM. \textit{JLSRF}, \textbf{2}, A60 (2016).\\

\end{document}